\documentclass[preprint,aps,tightenlines,floatfix,showpacs]{revtex4}

\usepackage{graphicx}
\usepackage{bm}
\usepackage{amsmath}
\usepackage{amssymb}

\begin{document}
\title{Constraints on the spectra of $^{17,19}$C}

\author{S. Karataglidis$^{(1)}$}
\email{S.Karataglidis@ru.ac.za}
\author{K. Amos$^{(2)}$}
\email{amos@physics.unimelb.edu.au}
\author{P. Fraser$^{(2)}$}
\email{pfraser@physics.unimelb.edu.au}
\author{L. Canton$^{(3)}$}
\email{luciano.canton@pd.infn.it}
\author{J. P. Svenne$^{(4)}$}
\email{svenne@physics.umanitoba.ca}

\affiliation{$^{(1)}$Department of Physics and Electronics, Rhodes
University, P.O. Box 94, Grahamstown 6140, South Africa}
\affiliation{$^{(2)}$School of Physics, University of Melbourne,
Victoria 3010, Australia}
\affiliation{$^{(3)}$Istituto Nazionale di Fisica Nucleare, sezione di
Padova,\\
e Dipartimento di Fisica dell'Universit$\grave {\rm a}$
di Padova, via Marzolo 8, Padova I-35131, Italia}
\affiliation{$^{(4)}$
Department of Physics and Astronomy, University of Manitoba,
and Winnipeg Institute for Theoretical Physics,
Winnipeg, Manitoba, Canada, R3T 2N2}

\date{\today}

\begin{abstract}
Diverse means are used to investigate the spectra of the radioactive, 
exotic ions, $^{17,19}$C. First, estimates have been made using a 
shell model for the systems. Information from those shell model studies 
were then used in evaluating cross sections of the scattering of $70A$~MeV 
$^{17,19}$C ions from hydrogen. Complementing those studies, a multichannel 
algebraic scattering (MCAS) theory for $n$+$^{16,18}$C coupled-channel 
problems has been used to identify structures of the compound systems. The 
results show that the shell model structure assumed for these ions is 
reasonable with little need of effective charges. The conditions that
two excited states exist within a few hundred keV 
of the ground state places some restriction upon the structure models. 
Other positive parity states are expected in the low-lying spectra
of the two nuclei.
\end{abstract}

\pacs{21.10.Hw,25.30.Dh,25.40.Ep,25.80.Ek}

\maketitle

\section{Introduction}

The properties of nuclei at and near to the nucleon drip lines are the subject
of some current interest as, increasingly, exotic nuclei are being produced in
radioactive ion beams (RIB) with which scattering experiments can be made.
Particular interest lies with such nuclei that have large neutron  (or 
proton)  excess  distributed  in  the extended spatial manner termed a 
nucleon halo~\cite{Ha95}.  The neutron-rich carbon and oxygen isotopes 
are of special interest with the conjecture that variation of
deformation of their ground states point to a new magic number of 16.
Another conjecture is that the ground state deformation is neutron number
dependent. That is suggested by deformed Hartree-Fock studies~\cite{Su03}  
from which  the isotopes $^{15,16,17}$C are expected to have prolate 
deformation. However, with $^{19}$C, prolate and oblate deformations
were found to be almost degenerate.

The neutron-rich nuclei $^{17,19}$C are two exotic nuclear systems
that have received some attention recently. Of note is that they have been
used in RIB, at $70A$~MeV energy, and cross sections measured \cite{Sa08}
in scattering from hydrogen 
targets. While both nuclei have large neutron excess, $^{19}$C 
is of particular interest since it lies at the drip line, has a small single-neutron 
separation energy, $\sim 0.53$ MeV, and most likely can be 
classified as a 
one-neutron halo nucleus \cite{Na99}. However, the spin-parity of its ground state
is still uncertain.  A standard shell model (SM)~\cite{Ot01} gave a 
$\frac{1}{2}^+$ assignment making the nucleus a candidate to be a 
$1s_{\frac{1}{2}}$-neutron-halo nucleus.   This is also suggested 
from studies of the Coulomb dissociation of $^{19}$C  in
collisions with heavy mass targets (Ta and Pb)~\cite{Na99,Ba00}. The 
same conjecture regarding  the ground state, and suggestions of  the 
spin-parities of low-lying excited states, were given in the  recent 
study by Elekes \textit{et al.}~\cite{El05}.     Their predictions were 
based upon SM evaluations, though in their paper they 
show only a few of the possible states. We have made independently an
SM calculation to get a spectrum of $^{19}$C   using the 
code OXBASH~\cite{Ox86} with the WBP~\cite{Wa92} interactions    and 
within the $spsdpf$ model space.   From that calculation we found 
that there may be more than one candidate for the ground state.

An alternative approach has been used by Ridikas \textit{et al.} \cite{Ri97,Ri98} to
obtain the states in $^{17,19}$C as a neutron coupled to a core in a coupled-channels
model. Their results
suggested that $^{19}$C in its ground state qualified as a
neutron-halo and  would have either
a $\frac{1}{2}^+$ or a $\frac{3}{2}^+$  spin-parity assignment.
The $\frac{5}{2}^+$ option was deemed unlikely. They also
noted that coupled-channel effects were important in this
description. But there is a possible 
problem with that since assurance of the Pauli principle is not 
guaranteed with their approach.
Such assurance is feasible within the MCAS
scheme~\cite{Ca05} and how important that is has been
demonstrated in Refs.~\cite{Ca06,Ca06a,Pi05}.
We will show the extent of that problem
with $^{19}$C herein. Notably in this case blocking of the
$0s_{\frac{1}{2}}$, $0p_{\frac{3}{2}}$, $0p_{\frac{1}{2}}$, and
(at least partially) the $0d_{\frac{5}{2}}$ and $1s_{\frac{1}{2}}$--orbitals, which 
Ridikas \textit{et al.} \cite{Ri97,Ri98} have not ensured, has impact.

Nucleon-nucleus scattering at medium energies, which under inverse 
kinematics equates to
scattering of nuclei from hydrogen targets,   is an excellent means by
which the matter density of the nucleus may be studied.  Microscopic
models now exist with which predictions of data from both elastic  and
inelastic scattering reactions can be made. When good, and   detailed,
specification of the nucleonic structure of the nucleus is used, those
predictions usually agree very well with observation~\cite{Am00}, both
in shape and magnitude. To facilitate such analyses of data, one first
must specify the nucleon-nucleus ($NA$) interaction. To do so requires
two main ingredients:
a) an effective nucleon-nucleon ($NN$) interaction in-medium, allowing
for the mean field as well as Pauli-blocking effects; and
b) a credible model of structure of the nucleus that is nucleon-based.
The process of forming optical potentials in this way has been termed 
$g$-folding~\cite{Am00}.  
With relative motion wave functions defined using $g$-folding potentials, 
a distorted wave approximation (DWA) suffices to analyze most
inelastic scattering data.  With such conditions met, past studies of 
the scattering (elastic and inelastic) of helium and lithium isotopes 
from hydrogen targets~\cite{Am00,La01,St02} revealed that $^6$He and 
$^{11}$Li could be characterised as neutron halo nuclei while 
$^8$He and $^{9}$Li were not. However, the latter pair was seen to
have their neutron excess as a skin-like distribution. The
differential cross sections from both elastic and inelastic scattering 
of those ions from hydrogen had signatures of those excess neutron 
distributions.

In the next section, we specify details of the SM calculations
for $^{17,19}$C we have made, and compare the spectra with the
few states of these systems that are known. We also specify other
properties of these systems, such as root mean square radii. 
In Sec.~\ref{ccsect} we consider the nuclei as coupled-channel problems 
of a neutron on $^{16,18}$C, allowing coupling to the ground and
$2^+$ excited states in particular. The sensitivity to aspects of the
coupling interaction are delineated and the crucial effect of Pauli
blocking and hindrance with these coupled-channel studies is
highlighted. Then in Sec.~\ref{pscat} the results of DWA analyses
of the scattering of $70A$ MeV $^{17,19}$C ions from hydrogen
are compared with the recently available data~\cite{Sa08}. 
Conclusions are then given in Sec.~\ref{concl}. 

\section{The shell model and $^{17,19}$C}
\label{SMsect}
\begin{figure}[h]
  \scalebox{0.6}{\includegraphics*{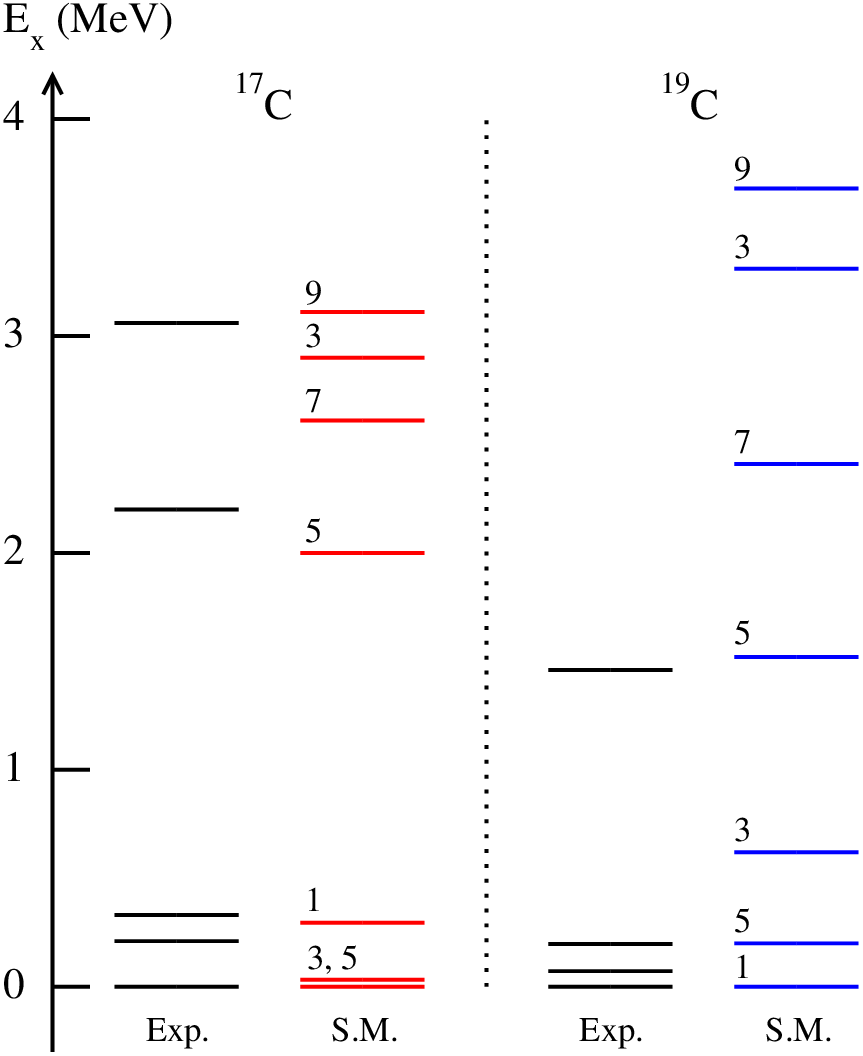}}
  \caption{\label{Fig1} (Color online.)
   The spectra of $^{17,19}$C generated from our SM calculations
  with the WBP potentials compared to experimental values.
  The number identifying each (positive parity) state is $2J$
  (twice the spin of the state).}
\end{figure}

The spectra of $^{17}$C and $^{19}$C, as obtained from the SM
calculations made using the WBP interaction~\cite{Wa92}, are displayed 
in Fig.~\ref{Fig1}. Only positive parity states are shown since
there is no knowledge to date of any states of negative parity. 
The shell-model spectrum of $^{17}$C has a ground state whose spin-parity
and isospin assignment ($J^\pi; T$) is $\frac{3}{2}^+;\frac{5}{2}$. Then  
the first excited state at 32~keV has spin-parity $\frac{5}{2}^+$.  
Note that a small perturbation in the matrix elements in the
interaction may invert these states.    The same ordering of states is
also observed in the spectrum obtained using the MK3W interaction, but
the first excited state is at 218~keV.       A low-energy spectrum for
$^{17}$C was measured by Elekes \textit{et al.}~\cite{El05},   wherein
two excited states were observed:  one at 210~keV and the other at 331
keV.    Those were given tentative assignments of $\frac{1}{2}^+$ and 
$\frac{5}{2}^+$
respectively,   compared to the ground state specification of 
$\frac{3}{2}^+$.
Our result confirms that of Ref.~\cite{El05} and which
they designate as the \textit{psdwbp} shell-model result.  
That was also
obtained using the WBP interaction.    Higher excited states have been
observed at 2.20, 3.05, and 6.13 MeV~\cite{Sa08}.   The states at 2.20
and  3.05  MeV  have  been  given  assignments of $\frac{7}{2}^+$ and 
$\frac{9}{2}^+$,
respectively,      and we find equivalent states at 2.61 and 3.11~MeV.
There is no evidence currently for excited states at 0.295 ($\frac{1}{2}^+$),   
2.00 ($\frac{5}{2}^+$), and 2.90 ($\frac{3}{2}^+$) MeV.

The observed spectrum of $^{19}$C~\cite{El05} has a  ground state
with $J^\pi;T  = \frac{1}{2}^+;\frac{7}{2}$. Then there
are  two  excited  states at 72 and 197~keV excitation.
Those were assigned spins of $\frac{3}{2}$ and $\frac{5}{2}$ based on 
the  assignments
to states in $^{17}$C.  A reversal of the spin assignments in $^{17}$C
would also lead to the reversal of those in $^{19}$C.   A
higher excited state has been observed at 1.46~MeV~\cite{Sa08},    for
which the assignment is $\frac{5}{2}^+$. Note that this is acknowledged 
as the
second $\frac{5}{2}^+$ state~\cite{Sa08}, consistent with our shell-model
predictions.

As a first test of the model of structure, one can evaluate 
various root mean square (rms) radii. With radioactive nuclear ions 
one can link their matter rms radii to the reaction
cross sections from their scattering off stable target nuclei~\cite{Oz01}.
In general, though, there are problems in doing so reliably as the usual 
methods of analysis, using Glauber theory of scattering and making 
additional approximations, can lead to disparate results. 
That is especially so if the ion in question
has weakly bound neutrons that give it an extended neutron matter 
distribution. Such was found for $^{19}$C in particular~\cite{To99}.
In that study, both the optical limit (OL) and  few-body (FB) approximations
were used to analyze the reaction cross sections of $^{19}$C ions
scattering from $^{12}$C. As shown 
in their Fig.~5, with a neutron separation energy of 0.5 MeV, the reaction
cross section data gave rms radii of $\sim 2.97$ and $\sim 3.12$~fm when
analyzed using the OL and FB schemes respectively.  Using Glauber theory
to analyze their interaction cross section data,
Ozawa \textit{et al}.~\cite{Oz01} found values of $\sim 2.72$~fm
for the matter rms radius of $^{17}$C and of $3.13$~fm (OL)
and $3.23$~fm (FB).
However, from an analysis of the reaction cross sections from the
scattering of $^{17,19}$C ions from $^{\rm nat}$Cu made using Karol's
prescription~\cite{Ka75}, Liatard \textit{et al.}~\cite{Li90} deduced
matter rms radii of $3.04$ and $2.74$~fm for the $^{17,19}$C ions respectively.
The latter results imply that $^{17}$C would have a more extended
neutron distribution than $^{19}$C.

In Table~\ref{radii} the experimental rms matter radii for 
$^{17}$C and $^{19}$C ions assessed from their interaction
cross sections from $^{12}$C targets, the FB and OL method results,
are compared with the 
pure SM results we have found and with a model prescription defined in
ref.~\cite{La92}. The SM results are given in the columns 
identified by `WBP' and were obtained using the WBP interaction and   
harmonic oscillator single particle wave functions ($b = 1.6$~fm)
to specify the shell-model Hamiltonians.
\begin{table}
\begin{ruledtabular}
\caption{\label{radii} Root mean square radii (in fm) for $^{17,19}$C.} 
\begin{tabular}{c|c|ccc}
Nucleus & $\sqrt{\left\langle r^2_C \right\rangle}$
\hspace*{0.5cm} 
& \multicolumn{3}{c}{$\sqrt{\left\langle
 r^2_T \right\rangle}$} \\
\hline
& WBP
\hspace*{0.5cm} 
& WBP & Ref.~\cite{La92} & Expt.~\cite{Oz01} \\
\hline
$^{17}$C & 2.425 
\hspace*{0.5cm} 
& 2.500 & 3.1 & $2.72 \pm 0.04$ \\
$^{19}$C & 2.422 
\hspace*{0.5cm} 
& 2.556 & 3.7 & $3.23 \pm 0.08$ (FB)\\
& & & & $3.13 \pm 0.07$ (OL) \\
\hline
$^{19}$C (halo) & 2.699 
\hspace*{0.5cm} 
& 3.982 & & \\
\end{tabular}
\end{ruledtabular}
\end{table}
The evaluated charge $\left(\sqrt{\left\langle r^2_C \right\rangle}\right)$ 
and matter $\left(\sqrt{\left\langle r^2_T \right\rangle}\right)$ radii 
are given.
Clearly the evaluated radii are too small whatever the experimental
values should be. But one must remember that heavy-ion collisions
are mostly peripheral and methods of extraction of the radii from such
are subject to some uncertainty aside from the quoted errors that
relate to data. Nonetheless it has been anticipated from studies of
neutron break-up reactions~\cite{Ba95}  that $^{19}$C is a one-neutron 
halo system.  In these studies  we can have a density profile that is 
neutron-halo-like by using Woods-Saxon (WS) single-nucleon bound-state wave 
functions.  The WS potential chosen 
was that used in the past for $^{12}$C,    but with the most important 
$sd$ (valence) orbits having weak binding energy (0.53 MeV \cite{Na99}).    
Similar
considerations~\cite{La01,St02} lead to a neutron halo description for 
$^6$He. Doing so with the SM ground state one-body-density-matrix-elements
(OBDME)~\cite{Am00}, gave an rms matter radius for $^{19}$C of 3.982~fm as
listed in the row of Table~\ref{radii} defined by ${}^{12}$C (halo).
The associated charge radius is slightly larger also.
A cluster model calculation \cite{La92}, based on the separation energy of 
the valence neutron, gave 3.1 and 3.7~fm for the matter radii of $^{17}$C 
and $^{19}$C, respectively. 
Those are listed in Table~\ref{radii} under the column
heading of Ref.~\cite{La92}.
Our halo result for the matter radius of $^{19}$C 
is consistent with that result, but both are much higher than the experimental 
values.

Recently~\cite{Am06}, reaction cross sections of $^{17,19}$C as
RIB scattering from
hydrogen were shown to be a good measure of spatial distributions of
neutron excess in many exotic nuclei. That was so because that scattering
probes more of the nuclear interior than does heavy ion collisions
and the method of analysis was based upon a credible, microscopic
reaction theory using realistic models of the structure of the ions. 
We have used the same approach, folding the relevant Melbourne
effective, medium modified, $NN$ interactions~\cite{Am00} with the 
structure details given by the shell model, to predict elastic
scattering cross sections for $70A$~MeV $^{17,19}$C ions from hydrogen. 
With oscillator single nucleon bound state
wave functions, those calculations gave 423 and 461~mb for the 
relevant reaction 
cross sections. Using the WS (halo creating) set of bound state
functions, the reaction cross section of the $^{19}$C scattering was
584~mb. The much higher value of the reaction cross section from the WS (halo)
model is consistent with that found for $^6$He, where such was needed to reproduce
the measured value \cite{La01}. A measurement of the reaction
cross section for the scattering of $^{19}$C from hydrogen would therefore be desirable.

The differential cross sections, themselves, are shown and discussed 
later in Sec.~\ref{pscat}.

\section{$^{17,19}$C as a coupled-channel problem of $n$+$^{16,18}$C}
\label{ccsect}

Herein we report evaluations of spectra for $^{17,19}$C that have
been defined from coupled-channel calculations of the nucleon-core, 
$n$+$^{16,18}$C systems, using the multi-channel algebraic scattering
(MCAS) method~\cite{Ca05,Am03,Ca07}. The method is given in detail 
in those references and so is not presented again.
Of great importance in the 
MCAS approach is that Pauli 
blocking or hindrance can be accommodated in the Hamiltonian~\cite{Ca06a}, even if
a collective model is used to define the coupling interactions. In the collective model, such can be
achieved by using an orthogonalizing pseudo-potential (OPP) 
scheme~\cite{Am03}; an
approach used in defining spectra
(bound states and resonances) for the mass-7 isobars~\cite{Ca06}
and of a nucleus just beyond the neutron drip-line,
$^{15}$F~\cite{Ca06a}.  However, while it is reasonable, with the
$^{16,18}$C nuclei, to take the 
$0s_{\frac{1}{2}}$, $0p_{\frac{3}{2}}$, and
$0p_{\frac{1}{2}}$--neutron orbits to be totally blocked, we find that
only partial blocking of the $0d_{\frac{5}{2}}$--neutron orbit is 
appropriate. Partial blocking, of the $0p_{\frac{1}{2}}$--orbit,  
was needed in an earlier 
study of the spectrum of mass-15 nuclei~\cite{Ca06a}. In these studies,
the OPP strengths used were $10^5$~MeV for the totally blocked states
(essentially $\infty$) but the neutron $0d_{\frac{5}{2}}$--orbit was
hindered by using just a few MeV for the OPP strength in that orbit.

The MCAS program, to date, uses a collective model prescription for
the input interaction potential matrix, $V_{cc'}(r)$; $c, c'$
representing
the sets of quantum numbers that define each unique interaction
channel~\cite{Am03}.  Deformation of that interaction 
is taken to second order in the expansion
scheme.  Details of the interaction forms have been published~\cite{Am03}. 
The approach is as yet phenomenological with the  
attendant uncertainties in selection of parameter values though 
a microscopic approach based upon shell model wave functions
and two-nucleon potentials is being developed. 
With stable target systems, usually there are many compound nucleus
levels, bound and resonant, against which the coupling interaction
can be tuned~\cite{Am03}. That is not the case with most exotic
nuclei,
at least currently.  Indeed with $^{17,19}$C there are few known
states in their spectra and the spin-parity assignments of those few
are still uncertain.  However, it does seem appropriate that both
nuclei should have three states, including the ground state, within
an excitation energy of a few hundred keV.  One can reasonably expect
those states to have spin-parity assignments of $\frac{1}{2}^+$, 
$\frac{3}{2}^+$, and  $\frac{5}{2}^+$, though in what order they occur
is unclear with ${}^{17}$C. With $^{19}$C we will demand that the
ground state has the $\frac{1}{2}^+$ assignment, in line with
recent arguments~\cite{Sa08} that this nucleus has a very
extended neutron distribution designated as a neutron halo.
Likewise, while there is a reasonably known low-excitation spectrum 
for $^{16}$C, not much is known of the spectrum of the target nucleus
$^{18}$C. 

Consequently the channels chosen for the coupling in the
$n$+$^{16,18}$C studies have been limited to the ground, $0^+$, 
and a $2^+$. Excitation energies of that $2^+$ state were taken
from the compilations~\cite{Ti95,Ti95a} as 1.77 and
1.62~MeV for $^{16,18}$C respectively.
There are three other states known in
$^{16}$C lying below the $n$+$^{15}$C threshold with spin-parities
thought to be $0^+$, $2^+$, and $4^+$ forming a triplet with average
centroid energy $\sim 3.4$~MeV. This is reminiscent of a vibrator
spectrum with quadrupole phonon energy of $\sim 1.7$~MeV and we
view the two nuclei as such. Thus the triplet of states in
$^{16}$C, and those we assume may be found eventually in
$^{18}$C, we consider as two quadrupole phonon triplets
which will couple weakly to the ground (vacuum) state. Hence
we have ignored them in finding solutions of the coupled equations. 
By so restricting the problems to be one of two channels, we
also relate to the model used by Ridikas \textit{et al}~\cite{Ri98}.
\begin{table}[h]
\begin{ruledtabular}
\caption{\label{MCASpars} Parameter values used in MCAS
evaluations of the spectra of $^{17,19}$C.} 
\begin{tabular}{cccccc}
Nucleus & $V_0$ (MeV) & $V_{ls}$ (MeV) & 
$ V_{ll}$ (MeV) & $V_{Is}$ (MeV) & $\beta_2$\\
\hline
${}^{17}$C & $-$40.0 & 10.0 & -2.5 & 2.0 & 0.3 \\
${}^{19}$C & $-$44.5 & 6.0 & 0.0 & 2.7 & 0.4 \\ 
\end{tabular}
\end{ruledtabular}
\end{table}
The MCAS calculations leading to the spectra depicted in 
Figs.~\ref{Fig2} and \ref{Fig4}
were made using the parameter set listed in Table~\ref{MCASpars}. The
geometries of the potentials were taken with $R = 2.9$~fm and 
$a = 0.8$~fm. These are not considered as either a unique or even the
best Hamiltonian potentials given the paucity of data against which
we have to check. But they suffice to meet the limited constraints posed 
and to indicate on what the results are most sensitive.
While only positive
parity states are known (or suggested) in the spectra, taking
the same interaction to define negative parity states,
resonances of that parity are then found with MCAS evaluations
to lie in this region of low-energy excitation.
However, as there is no experimental information
about any negative parity states in $^{17,19}$C, hereafter, 
in all figures and discussions,
we only display and describe the positive parity spectra resulting 
from our calculations.

\subsection{The spectra of $^{17}$C}

\begin{figure}[h]
\scalebox{0.8}{\includegraphics*{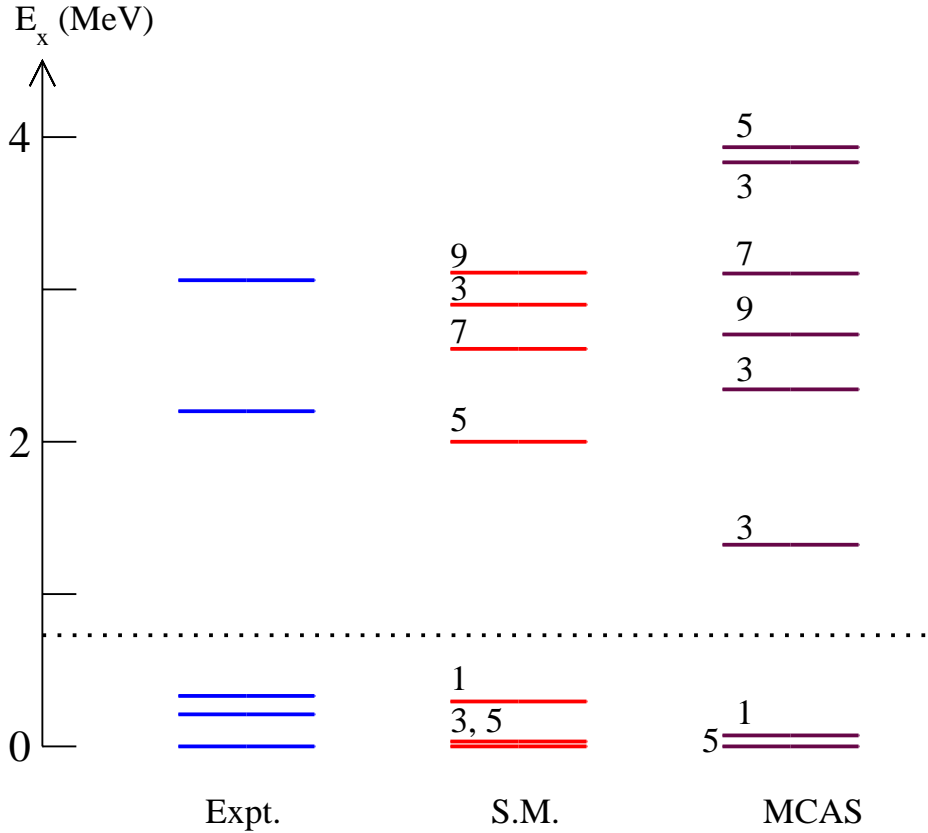}}
\caption{\label{Fig2} (Color online.)
The low excitation spectrum  of $^{17}$C. 
The data are compared with the results of a 
$2\hbar\omega$ SM calculation and
with a spectrum determined using the MCAS approach.
The dotted line indicates the $n$+$^{16}$C threshold.
The numeral identifying each state is again $2J$.}
\end{figure}

\begin{figure}[h]
\scalebox{0.8}{\includegraphics*{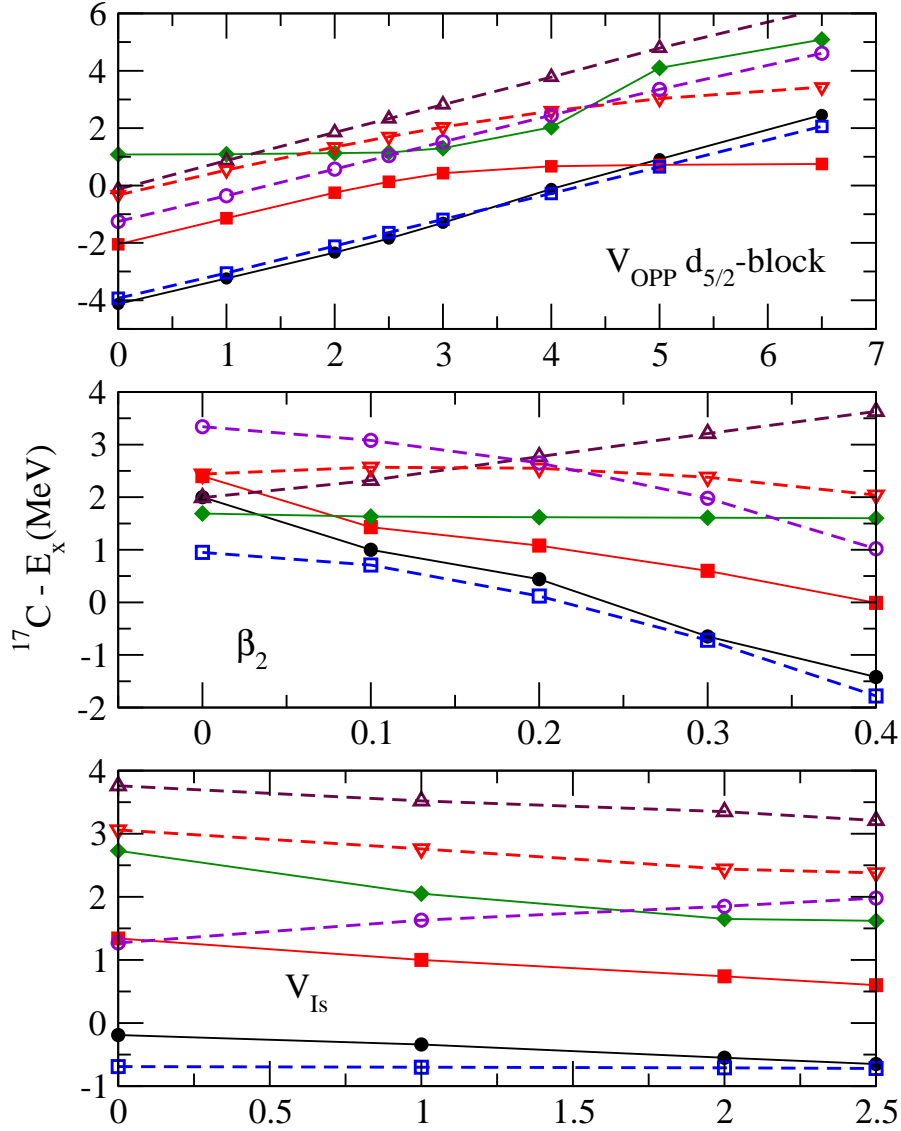}}
\caption{\label{Fig3} (Color online.) The low excitation spectrum  of
$^{17}$C from MCAS results found by varying the OPP for
$0d_{\frac{5}{2}}$-orbit blocking (units are MeV),
the deformation strength, $\beta_2$, and 
the spin-spin coupling, $V_{Is}$ (units are MeV) in the Hamiltonian. The various
lines are as described in the text.}
\end{figure}

\begin{figure}[h]
\scalebox{0.7}{\includegraphics*{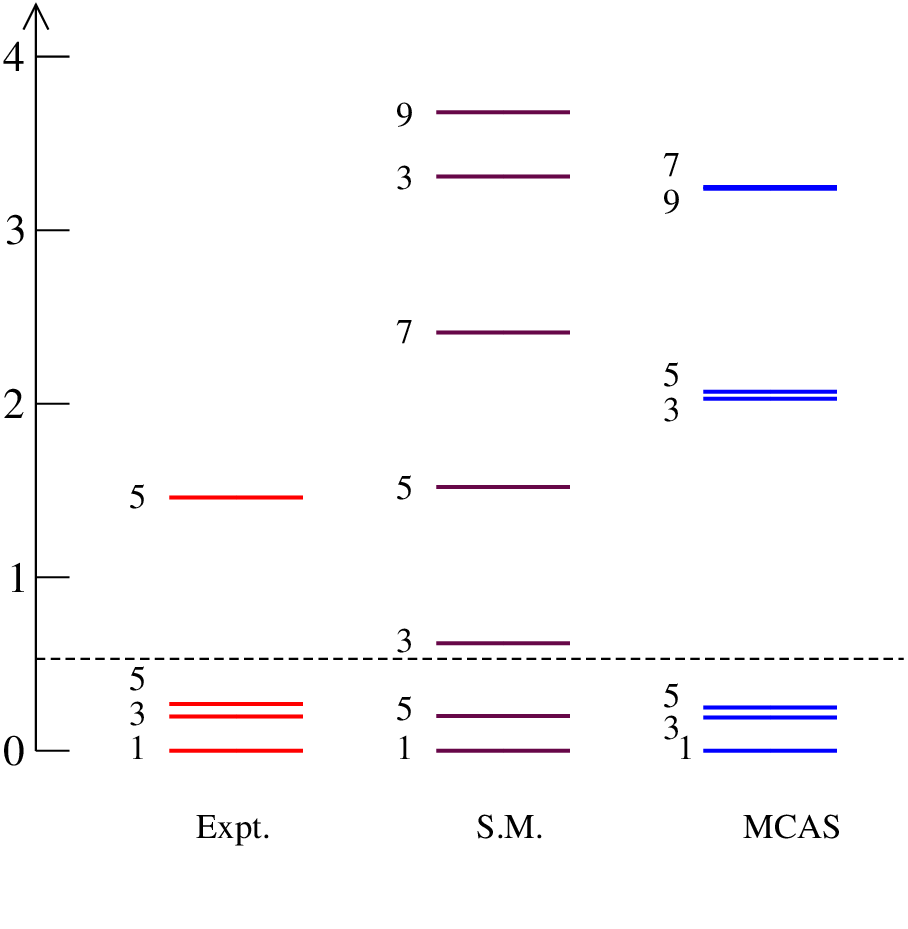}}
\caption{\label{Fig4} (Color online.) The low excitation spectrum  of
 $^{19}$C. The data~\cite{El05,Sa08} are compared with the results of a 
$2\hbar\omega$ SM calculation (SM) and
with a spectrum found using the MCAS approach.
The dashed line indicates the $n$+$^{18}$C threshold, while
the numeral identifying each state is again $2J$.}
\end{figure}

In Fig.~\ref{Fig2}, the currently known states in the spectrum of 
$^{17}$C  are compared with the low excitation spectra for this 
nucleus that have been determined from a shell-model calculation (SM) 
and from the coupled-channel solutions (MCAS).
The two theoretical calculations have very similar sets of 
states of spin-parity and both lead to two states within
$\sim 1$~MeV of the ground. The MCAS results are wider spread
in energy than those found with the shell model. That spread
persists despite the results being quite sensitive, in absolute
binding as well as state order, to the deformation, the amount of
$0d_{\frac{5}{2}}$-orbit blocking, and to the precise value
of the spin-spin coupling, $V_{Is}$, in the interaction form. 
To illustrate, we show in Fig.~\ref{Fig3}, the variations of
the spectra against each of those variables. We have not attempted
to adjust other parameters seeking to regain the absolute binding energy
of $\sim 0.7$~MeV.
In each panel, the individual states are associated with symbols
as follows: 
the $\frac{1}{2}^+$ state is shown by the filled circles,
the $\left. \frac{3}{2}^+ \right|_1$ state is shown by the filled squares,
the $\left. \frac{5}{2}^+ \right|_1$ state is shown by the open squares,
the $\left. \frac{3}{2}^+ \right|_2$ state is shown by the filled diamonds,
the $\frac{9}{2}^+$ state is shown by the open circles,
the $\frac{7}{2}^+$ state is shown by the open down triangles,
and the $\left. \frac{5}{2}^+ \right|_2$ state is shown by the open up triangles.

First consider the variations against the choice of the OPP strength
for Pauli hindrance of the $0d_{\frac{5}{2}}$ neutron orbit that are
displayed in the top section of this figure. By and large the states
simply track in parallel with increasing binding as the blocking is
decreased. However the $\frac{3}{2}^+$ states show a coupling effect, 
similar to orbit repulsion, in the 3 to 4 MeV strength range, to alter the
positions of those two states in the spectrum relative to others. 
 
In the middle panel of Fig~\ref{Fig3}, the variation of the
evaluated spectrum of $^{17}$C is shown as the deformation 
parameter, $\beta_2$, is changed. This variation of MCAS results
has been used previously~\cite{Pi05} to ascertain not only how the
channel coupling mixes basis states to optimally give a compound
nucleus spectrum, but also to delineate the basis state that is the 
progenitor of each compound system one. In the case of $^{17}$C
the two most bound states, having spin-parity of $\frac{5}{2}^+$ 
and $\frac{1}{2}^+$, remain close to each other whatever the
deformation value. With increasing
deformation, the other states in the spectrum spread apart.
 
Finally, in the bottom panel we display the variation of the spectrum 
with changes of the spin-spin, $V_{Is}$, term in the Hamiltonian.
Most states in the spectrum track more or less parallel in energy with 
increasing
interaction spin-spin strength. The exceptions are the
$\frac{9}{2}^+$ resonance state which increases
its centroid energy and the ground state whose energy hardly 
changes at all.

\subsection{The spectra of $^{19}$C}

In Fig.~\ref{Fig4}
the currently known states~\cite{El05,Sa08} in the spectrum of ${}^{19}$C  
are compared with the low excitation spectra for this nucleus
determined from a SM calculation and with that found
from the coupled-channel solutions (MCAS) based upon a two-state,
collective, model for the $n$+$^{18}$C system.  In this case,
the OPP strength of the neutron $0d_{\frac{5}{2}}$--orbit
was chosen as 100~MeV. Since that orbit is expected to have more neutrons 
contained than in $^{17}$C, variation of the $d_{\frac{5}{2}}$
blocking did not give as dramatic a change in the spectra, until
the strength of the OPP for that orbit was reduced to a value of 10 MeV 
or less. Then only the ground state ($\frac{1}{2}^+$) binding energy
was noticeably affected by being more bound by a few hundreds of keV.
However, taking the $d_{\frac{5}{2}}$ orbit to be Pauli allowed,
introduced spuriosity and the $\frac{1}{2}^+$  no longer is the ground state.
This is discussed more in the next section.

\begin{figure}[h]
\scalebox{0.7}{\includegraphics*{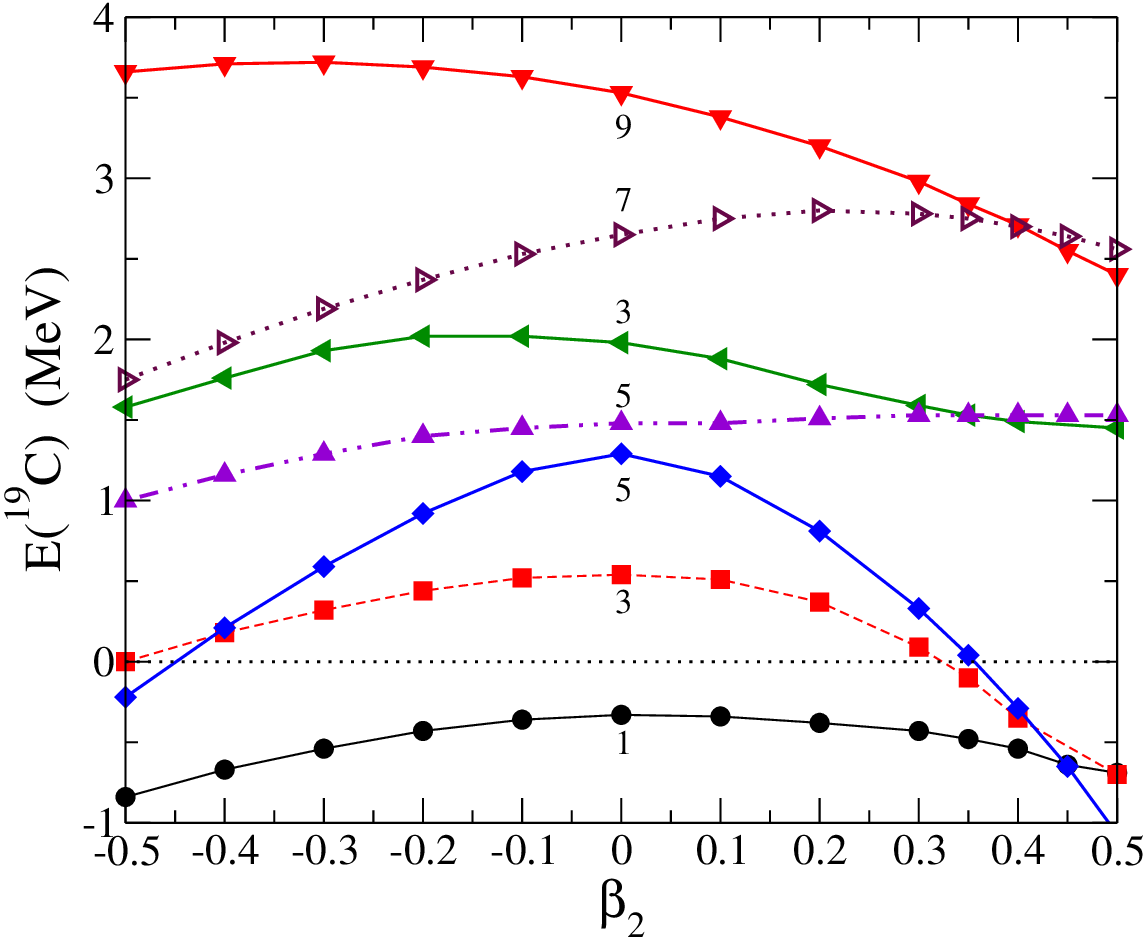}}
\caption{\label{Fig5} (Color online.) The low excitation spectrum  of
$^{19}$C calculated with variation of the deformation $\beta_2$.
The results for each spin-parity (in the order denoted in 
Fig.~\ref{Fig4}) are connected by lines drawn to
guide the eye.}
\end{figure}

The variation of the spectrum found by varying the deformation
parameter $\beta_2$ is displayed in Fig.~\ref{Fig5}. 
An equally dramatic, but quite different,  variation of results with $\beta_2$ was also
found in a study~\cite{Pi05} of the spectrum of ${}^{13}$C treated as
a coupled-channel, $n$+$^{12}$C, system. 
  The excitation energies of the states all vary regularly 
as the deformation increases, either for prolate (positive $\beta_2$)
or oblate (negative $\beta_2$) character. Two states stand out
as being dominantly the coupling of a single neutron to 
the ground state of $^{18}$C. They are denoted by
the filled circles (lowest set) being the ground state of $^{19}$C
formed (when $\beta_2 = 0$) with a $1s_{\frac{1}{2}}$--neutron, 
and by the filled triangles being a state formed (when $\beta_2 = 0$) 
with a $\frac{5}{2}^+$--neutron. These states vary with deformation
noticeably
more slowly than the others. That is so especially for the ground
state reflecting that the prime admixing component (the $\left. \left[ 0d_{\frac{5}{2}}
\otimes 2^+ \right]\right|_{\frac{1}{2}}$) lies above 4 MeV in the unperturbed   
spectrum.  But it is important to note that the deformation coupling 
mixes all basic states of given spin-parity to form the resultant ones
in the spectrum of $^{19}$C. Thus Ridikas \textit{et al.}~\cite{Ri98}
ascertained that the ground state could have as much as 25\% 
mixing of $0d \otimes 2^+$ with the basic $1s \otimes 0^+_{\text{g.s.}}$ 
component.   From this plot it is also clear that our coupled-channel
calculations require a strong deformation $\sim 0.4$ to obtain
three states of the appropriate spin-parity lying below the 
neutron-$^{18}$C threshold and still retaining a one-neutron
separation energy of $\sim 0.53$~MeV. 

The resonance states, found using MCAS and the deformation $\beta_2 = 0.4$,
 have neutron widths of 0.35~MeV
$\left( \left. \frac{3}{2}^+ \right|_2 \right)$, 2.6~MeV 
$\left( \left. \frac{5}{2}^+ \right|_2 \right)$, 0.47~MeV
$\left( \frac{9}{2}^+ \right)$, and of 0.71~MeV 
$\left( \frac{7}{2}^+ \right)$. If these widths
are credible, then it is more likely that the cross section measured 
by Satou \textit{et al.}~\cite{Sa08} in the scattering of $70A$~MeV
$^{19}$C ions from hydrogen is the excitation of the
$\left. \frac{5}{2}^+ \right|_2$ state. Structure models then have to find
that state $\sim 1$~MeV lower in energy.

The strength of the spin-spin interaction,
$V_{I\cdot s}$, is non-zero so that, when $\beta_2 = 0$, there is
a splitting of the degeneracy of states basically formed by a particle
coupling to the $2_1^+$ state in $^{18}$C.
Setting both $\beta_2$ and the strength of the spin-spin $\left( V_{I\cdot s} \right)$ to 
zero, gave the seven listed states but there are just four distinct state 
energies in the range. This limit gives the degeneracies that indicate the
primary coupling of each listed state. Of course, with 
deformation, the eigenvalues of the Hamiltonian are admixtures of 
the primary states of given $J^\pi$. 
The four resultant energies identify states ($nl_j \otimes {}^{18}$C) that are, 
with increasing energy, $1s_{\frac{1}{2}} \otimes 0^+_{\text{g.s.}}$, the doublet
$1s_{\frac{1}{2}} \otimes 2_1^+$, a $0d_\frac{5}{2} \otimes 0^+_{\text{g.s.}}$,
and a set that may originate from $0d_\frac{5}{2} \otimes 2_1^+$. 

\subsection{Pauli blocking}

\begin{table}[h]
\caption{\label{NoPP-list} 
Positive parity
states in $^{19}$C formed when the Pauli principle is not enforced.}
\begin{ruledtabular}
\begin{tabular}{ccccc}
\hline
$J^{(+)}$ & $\beta_2 = 0.4$ & $\beta_2 =0$ 
& $\beta_2 = V_{ss} = 0$ & Origin \\
\hline
$\frac{3}{2}^+$ & $-$22.9 & $-$22.9 & $-$19.9 & 
$0s_{\frac{1}{2}} + 2^+_1$ \\
$\frac{1}{2}^+$ & $-$21.7 & $-$21.5 & $-$21.5 & 
$0s_{\frac{1}{2}} + 0^+_1$\\
$\frac{5}{2}^+$ & $-$18.1 & $-$18.2 & $-$19.9 & 
$0s_{\frac{1}{2}} + 2^+_1$\\
$\frac{1}{2}^+$ & $-$2.02 & $-$0.33 & $-$0.33 & 
$1s_{\frac{1}{2}} + 0^+_1$\\
$\frac{5}{2}^+$ & $-$0.87 & 1.29 &  1.29 & 
$0d_{\frac{5}{2}} + 0^+_1$\\
$\frac{3}{2}^+$ &  1.45 & 1.97 &  2.82 & 
$0d_{\frac{5}{2}} + 2^+_1$\\
$\frac{5}{2}^+$ &  1.63 & 1.46 &  2.82 & 
$0d_{\frac{5}{2}} + 2^+_1$\\
$\frac{1}{2}^+$ &  0.77 &  1.83 & 2.82 & $0d_{\frac{5}{2}} + 2^+_1$\\
$\frac{9}{2}^+$ &  2.39 & 3.53 &  2.82 & $0d_{\frac{5}{2}} + 2^+_1$\\
$\frac{7}{2}^+$ &  2.56 & 2.65 &  2.82 & $0d_{\frac{5}{2}} + 2^+_1$\\
\end{tabular}
\end{ruledtabular}
\end{table}

We have shown in the foregoing discussion how changes occur in MCAS
results when, using the OPP method, Pauli hindrance of the
$0d_{\frac{5}{2}}$-orbit was made. In all of those analyses, the lower-lying 
single-particle orbits were taken to be Pauli blocked.
With a collective model of interactions, the coupled-channel approach to 
define the spectrum of an $(n+A)$ compound nucleus, 
we have shown previously~\cite{Ca05,Ca06,Ca06a,Pi05} 
how crucial it is to ensure that the Pauli principle is not violated
in the analysis.  We demonstrate how significant doing so is in one of the 
present cases. 
By not using the OPP in an MCAS evaluation of $n$+$^{18}$C, we find a plethora of
states in the spectrum.  The spectrum that results on using the best 
interaction potential (found for the problem with the OPP included)
and then setting all of the  OPP strengths to zero, is listed in 
Table~\ref{NoPP-list}. 
In this solution of the coupled-channel problem, without taking
the Pauli principle into account, 
the two most bound $\frac{1}{2}^+$ states and the most bound
$\frac{3}{2}^+$ and $\frac{5}{2}^+$ states are spurious. Even those
other levels which might be associated with actual states of the system, 
would have spurious components in their wave functions~\cite{Ca05,Pi05,Am05}.
A similar large set  of spurious negative parity states were found
when the negative parity interaction was taken to be the same as that
for the positive parity study.


\section{Elastic and inelastic scattering of $70A$ MeV ions}
\label{pscat}

We  have used a $g$-folding model to specify the optical
potentials with which predictions of elastic scattering observables have
been made.  Those potentials then have been used to find
the distorted waves that we  use in DWA calculations of inelastic
scattering. The effective two-nucleon interaction used in the  
$g$-folding was also used as the transition operator effecting the
inelastic transitions.  Full details are given in Ref.~\cite{Am00} 
and so are not repeated herein.


\subsection{The elastic scattering}

\begin{figure}[h]
  \scalebox{0.8}{\includegraphics*{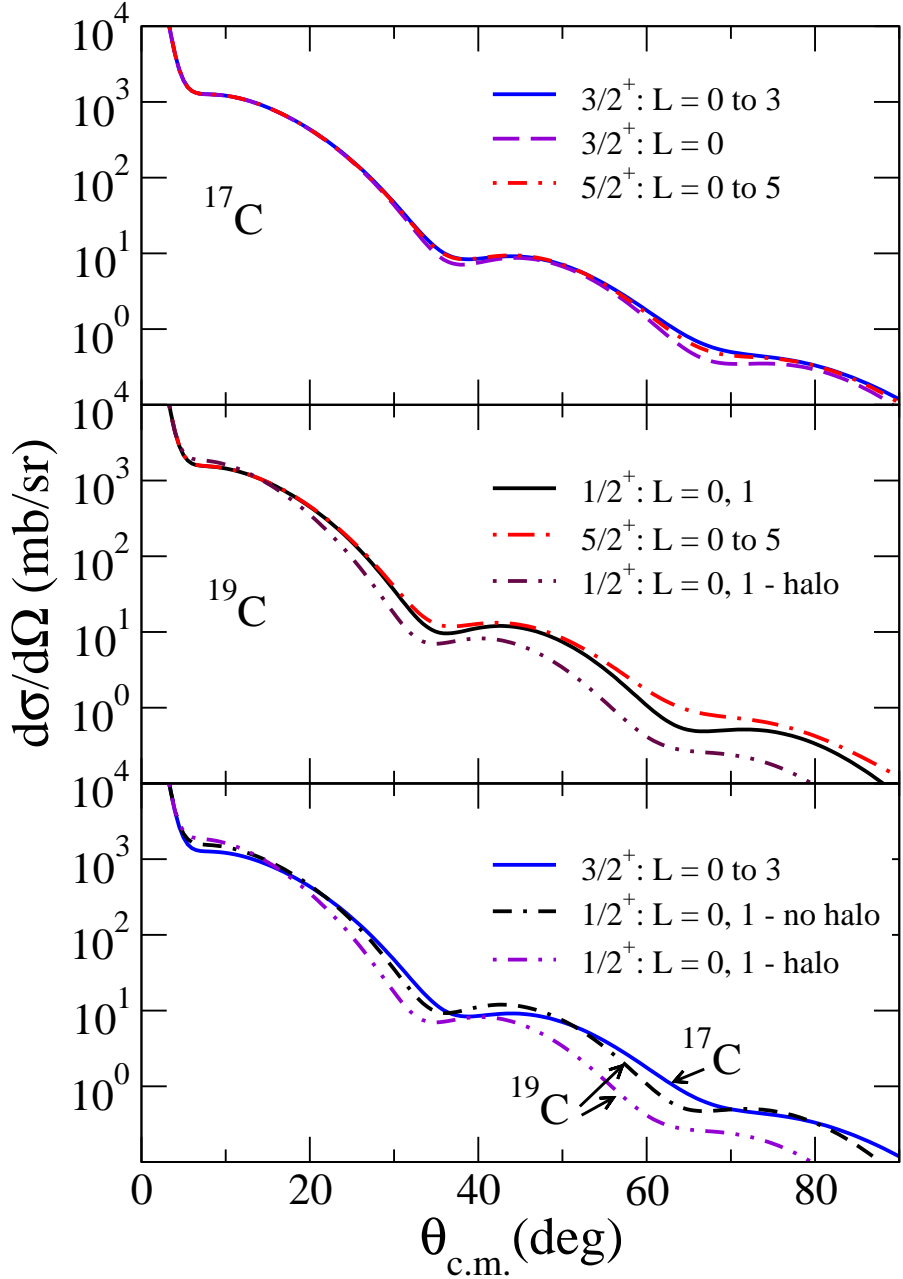}}
\caption{\label{Fig6} (Color online.)       The differential 
cross sections for the elastic scattering of $70A$~MeV  $^{17,19}$C ions 
from hydrogen.  The legends identify what spin-parity has been assumed 
for the ground states, and what transition multipoles have been taken into
account.
More details are given in the text.}
\end{figure}
Our $g$-folding model results for the elastic scattering cross 
sections (the kinematic inverse equivalent of $70A$
MeV    $^{17,19}$C    ions    from    hydrogen)    are    shown     in 
Fig.~\ref{Fig6}.     
As indicated,  the top set of results are for $^{17}$C 
scattering, the middle set are for $^{19}$C, and in the bottom panel we 
compare the results for the two nuclei.  In the top panel, there are 
three results shown for $^{17}$C. Of those, the solid curve is the total 
result  found  assuming  that  the  ground  state  has  spin-parity of 
$\frac{3}{2}^+$.  A number of multipoles ($L = 0, 1, 2, 3$) contribute 
to the scattering,  and  while  the  monopole term is defined from the 
$g$-folding optical potential,  the other multipole contributions have 
been found using a DWA method~\cite{Am00}.    In all though, the input 
information (OBDME and single nucleon wave functions, oscillators with 
$b = 1.6$ fm) was obtained from our SM calculations made using
the WBP 
potentials.   The dashed curve is the result when only the $L = 0$ 
components in that full calculation are retained.  When the (monopole) 
OBDME  determined  by  using  the  MK3W  potentials  in  a shell-model 
calculation are used, the cross section can hardly be    distinguished 
from this dashed curve. The dot-dashed curve is the total result found 
using OBDME from a SM calculation made with the WBP potentials    but 
assuming 
that the ground state has the $\frac{5}{2}^+$ specification.  There is 
very little difference among all of these results indicating that     the 
monopole terms dominate; that the choice of SM potentials
is not discernible (in the elastic scattering cross section); and that
the results do not show a preference for either the $\frac{3}{2}^+$ or 
$\frac{5}{2}^+$ choice of spin-parity of the ground state.

In the middle panel are the results for the elastic scattering of $70A$ 
MeV $^{19}$C ions from hydrogen which were also found using the  WBP SM 
information. The solid curve depicts the result found assuming  that  
the ground state has a spin-parity assignment of $\frac{1}{2}^+$.  
Almost the same result was found when the ground state was taken 
to have a spin-parity assignment of $\frac{5}{2}^+$  though only if a 
monopole interaction contribution is considered.  Allowing for all multipole 
contributions with a $\frac{5}{2}^+$ assignment, with amplitudes for 
non-zero  multipoles 
obtained using the DWA,  lead  to  the result shown by the dot-dashed 
curve.   There is some difference according to 
what is assumed for the ground-state spin-parity in this case,    more 
than seen with $^{17}$C,  but the differences overall remain small and 
it may be too hard to discern experimentally. Again, the single-nucleon 
wave functions used were harmonic oscillators with $b = 1.6$ fm. To consider 
whether $^{19}$C is a halo, we have also made evaluations with the single-nucleon wave 
functions being those of the WS potential specified earlier.  The
neutron density formed with such wave functions is quite extended; to
form what has been termed a neutron halo. 
Of course, even with oscillator wave functions, these nuclei 
($^{17,19}$C) have neutron skins. However, on using the WS functions 
in the $g$-folding process,   optical potentials result from which the 
cross    section    depicted    by    the    dot-dot-dashed  curve  in 
the middle panel of Fig.~\ref{Fig6}.  It differs sufficiently from the 
cross section found using the oscillator wave functions
that measurement should reflect which,  if either,    is the more
appropriate. Regrettably, no data for this elastic scattering have been
taken. 

Finally, in the bottom panel, we compare the cross sections we have 
evaluated for the two nuclei. 
Assuming that the spin-parities of the ground states are $\frac{3}{2}^+$
and $\frac{1}{2}^+$ for $^{17,19}$C respectively,
the solid curve is the $^{17}$C scattering cross section.
The other two are 
the $^{19}$C cross sections found using oscillator wave functions 
(dot-dash-dash curve) and the WS ones that form a neutron halo (dot-dot-dashed
curve).  The differences, especially if $^{19}$C
has a neutron halo and $^{17}$C does not, should be observable.

\subsection{The inelastic scattering}

         Satou {\it et al.}~\cite{Sa08} report data from the inelastic 
scattering of $^{17,19}$C ions from a hydrogen target.    Differential
cross sections from the scattering of $70A$ MeV ions leading to states 
at excitation energies of 2.2 and 3.05 MeV in $^{17}$C and to the 1.46 
MeV excited state in $^{19}$C were presented.     The analyses made in 
Ref.~\cite{Sa08} suggest that the ground states  have spin-parities of 
$\frac{3}{2}^+$ ($^{17}$C) and $\frac{1}{2}^+$ ($^{19}$C).       Those 
authors  also  suggest  that  the  excited  states  of  $^{17}$C  have 
spin-parity    assignments    of    $\frac{7}{2}^+$    (2.2 MeV)   and 
$\frac{9}{2}^+$ (3.05 MeV) while the 1.46 MeV state in $^{19}$C  has a 
$\frac{5}{2}^+$ specification.
But  it is unfortunate that no elastic scattering data have been taken, 
as replication of such by  predictions is a first level of  confidence 
as  to  the  structure  and  single-nucleon  wave  functions that best 
describe the radioactive ion.         

Past studies~\cite{La01,St02} of 
scattering of the halo nucleus, $^6$He, from hydrogen showed that  the 
extended neutron distribution for this ion,     had a marked 
effect upon the elastic as well as the inelastic scattering cross sections.
We 
proceed   under   the   assumption   that   the   results   shown   in 
Fig.~\ref{Fig6} may replicate future data.

\begin{figure}[h]
  \scalebox{0.8}{\includegraphics*{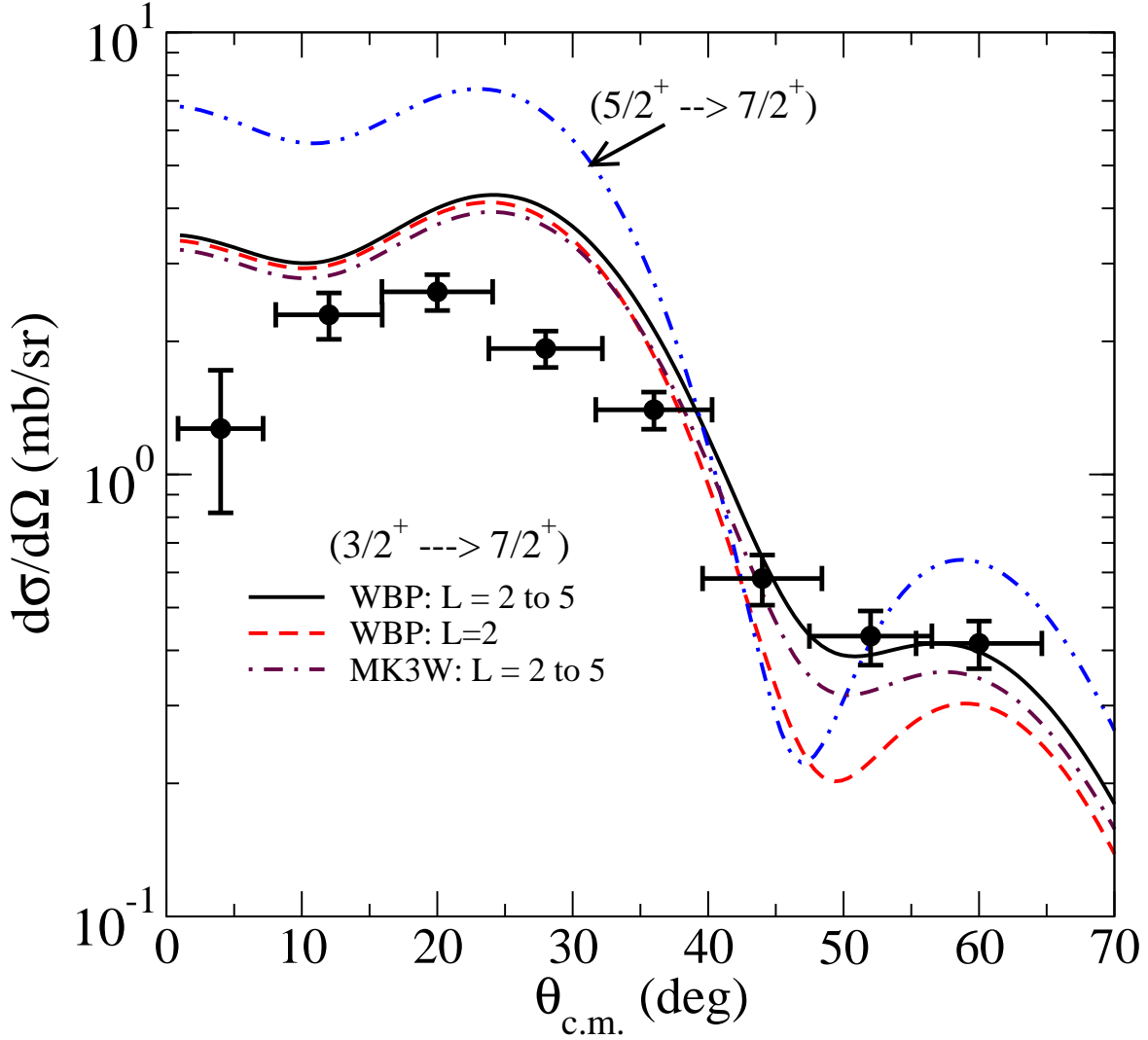}}
\caption{\label{Fig7} (Color online.)    The differential cross
sections for the inelastic scattering of  70 MeV protons from $^{17}$C
leading to the state at 2.2 MeV excitation. What  
has been assumed to be the spin-parities of the initial and final states
in the transition are indicated in the brackets, while for the assumed
$\frac{3}{2}^+ \to \frac{7}{2}^+$ case, the legends indicate what shell
model was used and what multipoles considered in the transition. 
More details are given in the 
text.}
\end{figure}

 First we consider the inelastic scattering of $70A$ MeV $^{17}$C ions
from  hydrogen  leading  to a state  in the radioactive ion at 2.2 MeV 
excitation.      Data~\cite{Sa08} for this excitation are displayed in
Fig.~\ref{Fig7}.
Assuming that the spin-parity of the excited state is  $\frac{7}{2}^+$
the DWA results in best agreement with this data were based upon   the 
ground state having a spin-parity assignment of $\frac{3}{2}^+$.   The 
complete result is depicted by the solid curve.    The dashed curve is 
the result of using only the    $L=2$    transfer contributions in the 
transition.   Clearly the $L=2$ transition dominates, and is 
in quite reasonable agreement with the observation. The third curve of 
this set, shown as the dot-dashed curve,     is the cross section that 
results using the pure  $L=2$  transition  OBDME obtained from the
shell -model calculation made using the  
MK3W interaction.   The complete result 
found using that (MK3W) shell-model OBDME,  is so similar to the solid 
curve that it is not shown. The fourth curve in Fig.~\ref{Fig7}, 
the dot-dot-dashed curve,    portrays the cross section found when the 
ground state is assumed to have $\frac{5}{2}^+$ spin-parity assignment. 
Comparisons   with   data   clearly   suggest   the $\frac{3}{2}^+$ 
spin-parity assignment for the ground state.

\begin{figure}[h]
  \scalebox{0.8}{\includegraphics*{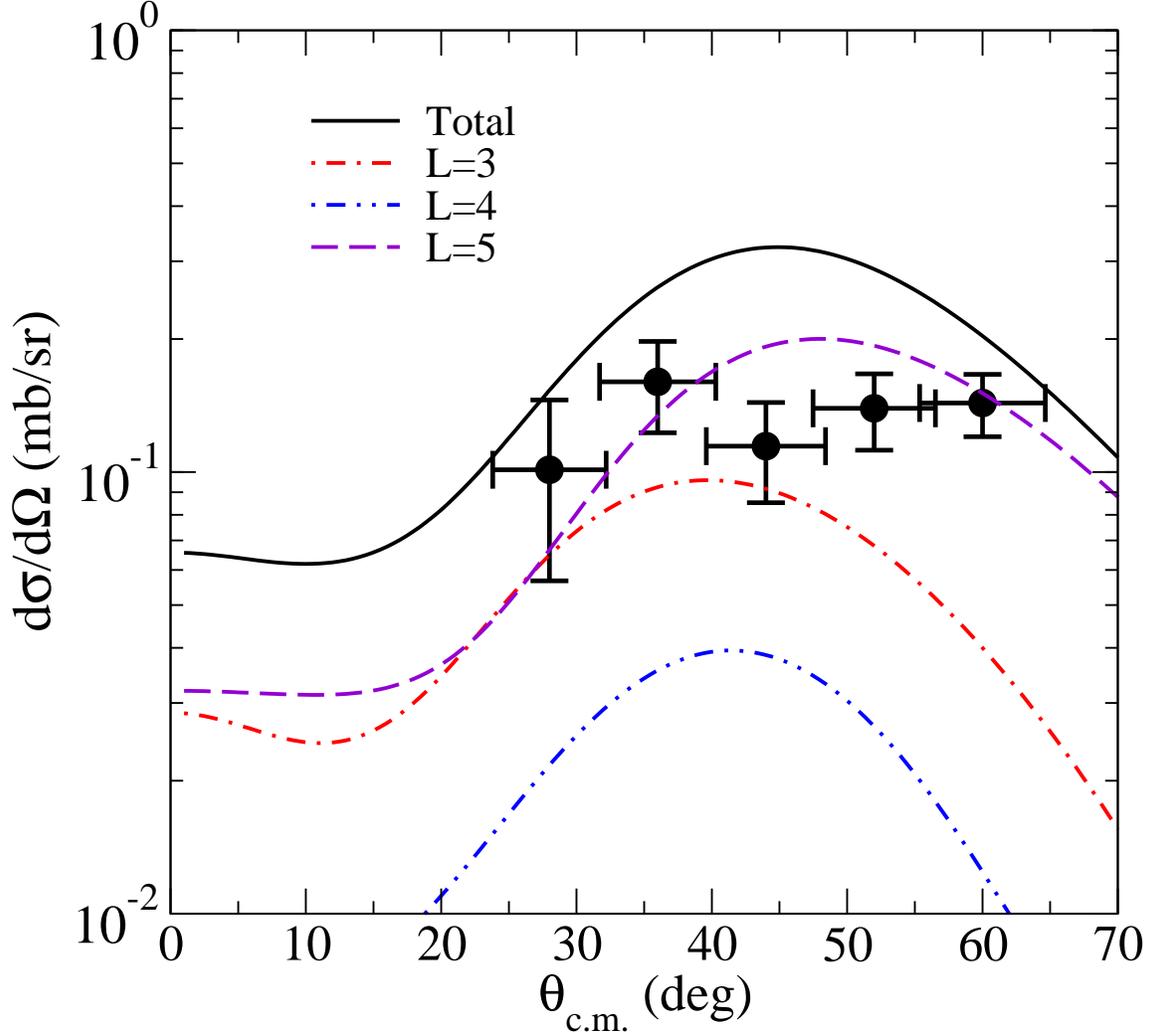}}
\caption{\label{Fig8} (Color online.)   The differential cross
sections for the inelastic scattering of  70 MeV protons from $^{17}$C
leading to the state at 3.05 MeV excitation. 
The legends identify which multipoles of the transition have been
kept in the calculation More details are given in the text.}
\end{figure}

     The state at 3.05 MeV excitation in $^{17}$C is assumed to have a
spin-parity of $\frac{9}{2}^+$,        and cross-section data from its 
excitation in the collision of  $70A$  MeV $^{17}$C ions with hydrogen
\cite{Sa08} are displayed in Fig.~\ref{Fig8}.
  Assuming that the ground state has a spin-parity of $\frac{3}{2}^+$,
then transition multipoles ($L$) of 3, 4, and 5 are possible.      The 
angular-momentum allowed case of $L = 6$ is not feasible given     the 
single-particle basis used in the shell-model calculations.  The total 
result is depicted by the solid curve in Fig.~\ref{Fig8}.  The 
separate component cross sections are portrayed by the dot-dashed, the
dot-dot-dashed, and the long-dashed curves for  the $L = 3, 4$, and 
5 contributions respectively.   Clearly our prediction is slightly too 
large in comparison with the data, and it is the $L = 5$    associated 
with recoupling of a $d_{\frac{5}{2}}$ neutron orbit that is primarily 
responsible.

\begin{figure}
  \scalebox{0.8}{\includegraphics*{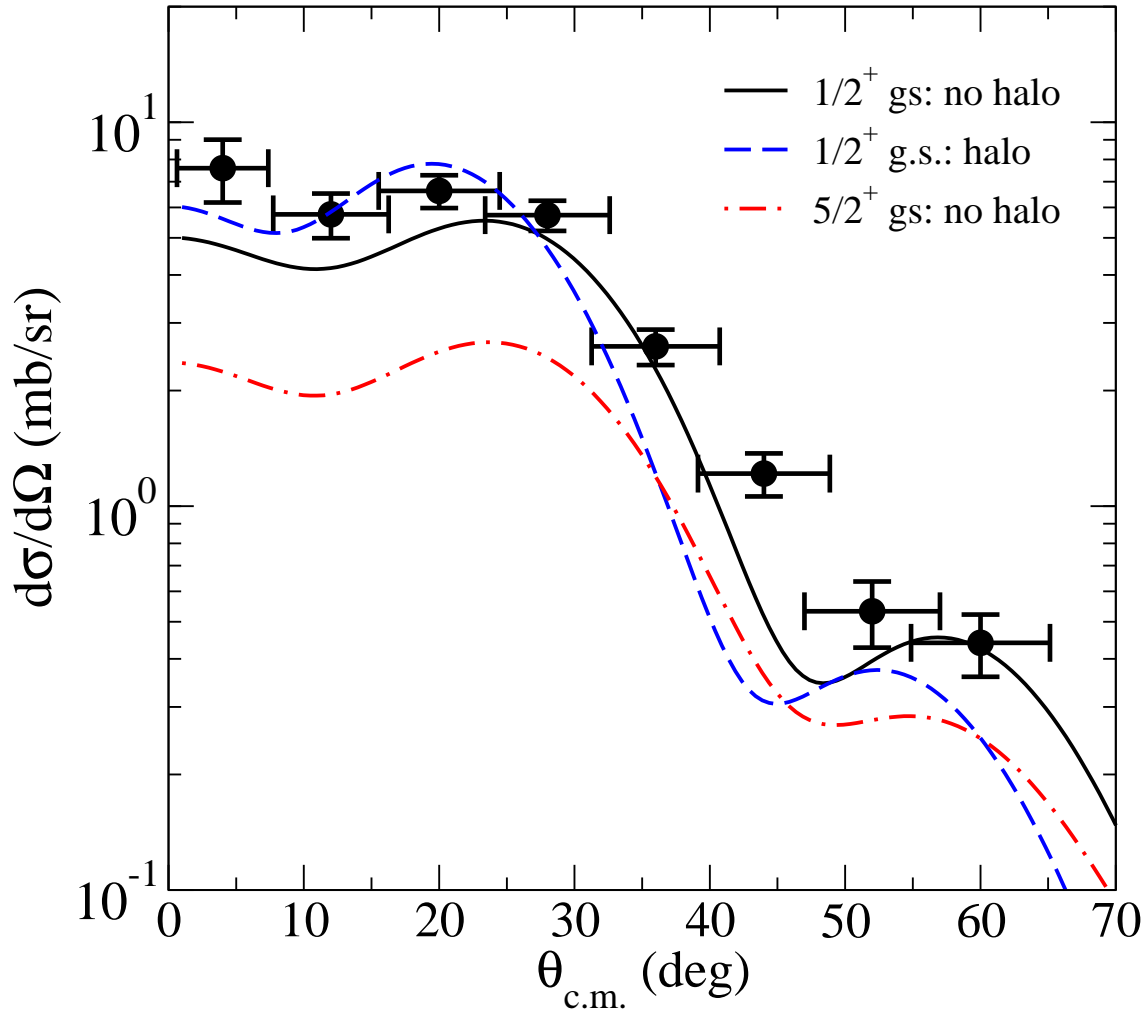}}
\caption{\label{Fig9} (Color online.)   The differential cross
sections for the inelastic scattering of  70 MeV protons from $^{19}$C
leading to the state at 1.46 MeV excitation.  
The legends indicate what has been assumed to be the spin-parity
of the ground state and whether oscillator (no halo) or the Woods-Saxon
(halo) functions were used.  More details are given in the text.}
\end{figure}

Satou {\it et al.}~\cite{Sa08} also measured the cross section for the 
excitation of the 1.46~MeV state 
in $^{19}$C from the scattering of that ion from hydrogen. 
Those data are shown in Fig.~\ref{Fig9}.
There are three evaluations with which these data are compared.  Those 
depicted by the solid and  long-dashed  curves were made assuming that 
the ground state had a spin-parity of $\frac{1}{2}^+$.    Both results 
are dominated by $L=2$ angular momentum transfer.   The solid (dashed)
curve is the result found using oscillator (Woods-Saxon) functions for
the single-nucleon bound-state wave functions that reflect  a  neutron
skin (halo-like) property to the density.   The result depicted by the
dot-dashed curve, was found using the oscillator wave functions but on
assuming  that  the ground state had a spin-parity of $\frac{5}{2}^+$. 
These results indicate that the ground state of $^{19}$C has indeed the 
$\frac{1}{2}^+$ assignment as has been suggested~\cite{Sa08}, but that 
it may have a neutron skin rather than a neutron halo.          Again, 
measurement of  the  elastic  scattering  cross  section  may  give  a 
resolution  between  the alternatives for the extended distribution of 
neutrons in this ion.

\section{Conclusions}
\label{concl}

These studies of the spectra of ${}^{17,19}$C, made with quite
complementary 
models (SM and MCAS) for their structure, reveal that they are very
sensitive to details used in these models. This presents a very clear
need for many more experimental studies of the systems to be made to
enumerate many more states in their spectra, at low excitations
especially, and to determine the spin-parity assignments most
appropriate for each state therein. Scattering data
are also needed to provide the best 
possible test of the wave functions deemed most suited to describe the ions.
In particular, cross sections from RIB elastic scattering  from
hydrogen targets, and eventually with electrons (as may be done using
the self confining radioactive ion targets (SCRIT)
system~\cite{Su05}), should be of most use given that the theories 
available to describe those reactions are formulated, currently, with 
least approximation problems.

\section*{Acknowledgments}
This research was supported by the Italian MIUR-PRIN Project
``Fisica Teorica del Nucleo e dei Sistemi a Pi\`u Corpi'', by the
Natural Sciences and Engineering Research Council (NSERC), Canada,
by the National Research Foundation of South Africa, and by
support from the International Exchange Program of the Australian
Academy of Science.

\bibliography{p+16and18C}

\end{document}